
\large

\begin{center}
{\bf Photon statistics of multimode even and odd coherent light }
\end{center}

\bigskip

\begin{center}
Nadeem A. Ansari and V.I. Man'ko\footnote{On leave from Lebedev Physics
Institue, Mascow, Russia.}
\end{center}

\bigskip

\begin{center}

Departimento di Scienze Fisiche, Universita` di Napoli "Federico II"
Istituto Nazionale di Fisiche Nucleare, Sezione di Napoli\\
Mostra d'Oltremare, Pad. 20-80125 Napoli, Italy
\end{center}

\bigskip

\begin{abstract}
The even and odd coherent states are generalized for multimode case.
The explicit forms for the photon distribution, Q-function and Wigner
 function are derived. In particular, it is shown that for two-mode
case there exist strong correlations between these modes, under certain
conditions, which are responsible for two-mode squeezing in case of even
coherent states
\end{abstract}

\newpage

\section{Introduction}

The even and odd coherent states of one mode oscillator have been introduced
in [1] and they are studied in detail in Refs.[2-4]. The theoretical
predictions of their possible generation  have been discussed in Refs.[5-9].
The connection of these states with `Schrodinger cat' states has been
analysed specially because of the non-classical behaviour of these states [9].
An important possible application of these states has been predicted
in interferometric detection of gravitational waves for reducing the optimal
intensity of the input laser[10]. The properties of even coherent states are
similar to the properties of squeezed vacuum states[11-14] and correlated
vacuum
states[15], because all such states are linear combination of the even photon
number states. The only difference lies in the appearance of the normalization
constants.These similarities allow us to investigate the possibilities to
replace the squeezed light with even coherent light in the situations when the
squeezed light shows some important applications.
This problem has been posed and
received a positive answer in Ref.[10], where the improvement for the
sensitivity  for the gravitational waves
interferometric detection is suggested to achieve
by using even coherent light as an alternative source to squeezed vacuum.
Also it was shown that even coherent states may play an important role
for enhancing the steady state squeezing in degenerate parametric
oscillator when they are coupled to the cavity from outside by
partially reflecting cavity end mirror [16].

Photon statistics of one-mode even and odd coherent states have been discussed
in Ref.[2,3], where the photon distribution function has been obtained and
nonclassical properties such as squeezing in quadrature component
and higher-order squeezing phenomena are considered.
Among the different theoretical predictions for the generation of such states,
Gerry [5], has predicted that such single-mode even and odd coherent states
can be produced in a Mach-Zender interferometer with Kerr medium and with the
interaction of two-level atoms with the input coherent light.
Recently, Gerry and Hach have shown another possibility of generating
even and odd coherent light from long-time evolution of the competition between
a two-photon absorption and two-photon parametric processes when the initial
fields state is considered to be  vacuum or one-photon state. Gea-Banacloche
[7], demonstrated the possibility for appearing such states in the resonant
Jaynes-Cummings model (see also [8]). Brune et al. [9] discussed the creation
of these states by dispersive atom-field couplings.

Till now the
study of even and odd coherent states has been concentrated on single-mode.
Such nonclassical light states corresponding to multimode field are also
interesting because they in principle, may be generated in the similar
situations which have been discussed for the single-mode case. Also the
multimode even and odd coherent states may be considered as
the simplest model of
"Schrodinger cat" states for many degrees of freedom.

The aim of this paper is to consider the multimode generalization of the even
and odd coherent states. We obtain the photon distribution function in
multimode even and odd coherent states and calculated means and matrix of
second
order momenta for photon numbers corresponding to these distributions. We have
shown that such distributions demonstrate essential differences with Poissonian
photon distributions for multimode coherent states.
Also we have evaluated the exact
expressions for the Wigner function and Q-function for
multimode even and odd coherent states. In addition we have studied the case of
two modes in detail and have derived the exact expressions for the photon
distribution function. We have also predicted the regions of different
parameters for two-mode
even and odd coherent states where the correlations between these two modes
develop between them and which causes the basis for two-mode squeezing.

\section {Single-mode even and odd coherent states}

In this section we will review the properties of single-mode even and odd
coherent states and on the ground of these calculations in the next section
we will develop the formalism for multi-mode even and odd coherent states.

Usually a coherent state $\mid \alpha>$ can be defined as [17]
\begin{equation}
\mid \alpha>=D(\alpha) \mid 0>,
\end{equation}
where the displacement operator D$(\alpha)$ is
\begin{equation}
D(\alpha)=exp[\alpha a^{\dag}-\alpha^{*} a].
\end{equation}
Also $\alpha$ is a complex number and $a(a^{\dag}$) being the annihilation
and creation operators for the coherent field. In addition to normalized
vacuum state $\mid 0>$, i.e., $<0 \mid 0>$=1, we have decomposition of the
normalized coherent states in terms of number states $\mid n>$ as
\begin{equation}
\mid \alpha>=e^{-\frac{\mid \alpha \mid^2}{2}} \sum_{n} \frac{\alpha^{n}}
{\sqrt{n!}} \mid n>,
\end{equation}
that corresponds to the photon distribution function for coherent states to be
the Poissonian distribution function[17,18]
\begin{equation}
P(n)=e^{-\mid \alpha \mid^{2}} \frac{\mid \alpha \mid^{2n}}{n!}.
\end{equation}
The coherent state $\mid \alpha>$ itself is the eigenstate of the photon
annihilation operator with eigenvalue $\alpha$, i.e.,
\begin{equation}
a \mid \alpha>=\alpha \mid \alpha>.
\end{equation}
The coherent states are the minimum uncertainty states. In Ref.[1,18]), it was
proposed to consider the superposition of two coherent states $\mid \alpha>$
and $\mid -\alpha>$. One of the possible superposition describes the even
coherent states
\begin{equation}
\mid \alpha_{+}>=N_{+}(\mid \alpha>+\mid -\alpha>),
\end{equation}
Where the normalization constant $N_{+}$ is of the form
\begin{equation}
N_{+}=\frac{e^{\frac{\mid \alpha \mid^2}{2}}}{2\sqrt{cosh
\mid \alpha \mid^{2}}}.
\end{equation}
For odd coherent state the following relation holds
\begin{equation}
\mid \alpha_{-}>=N_{-}(\mid \alpha>-\mid -\alpha>),
\end{equation}
and the normalization constant $N_{-}$ corresponding to the odd coherent states
is
\begin{equation}
N_{-}=\frac{e^{\frac{\mid \alpha \mid^2}{2}}}{2\sqrt{sinh
\mid \alpha \mid^{2}}}.
\end{equation}
 These states can be generated from vacuum in the following way
\begin{equation}
\mid \alpha_{\pm}>=D(\alpha_{\pm})\mid 0>,
\end{equation}
where displacement operators for the even and odd coherent states are related
to the displacement operator of the coherent states as these states
are the superposition of the coherent states and are following
\begin{eqnarray}
D(\alpha_{+})&=&cosh(\alpha a^{\dag}-\alpha^{*} a),\nonumber\\
D(\alpha_{-})&=&sinh(\alpha a^{\dag}-\alpha^{*} a).
\end{eqnarray}
Both states are the normalized eigenstates of the operator $a^{2}$
\begin{equation}
a^{2}\mid \alpha_{\pm}>=\alpha^{2}\mid \alpha_{\pm}>,
\end{equation}
where $\alpha$ is an arbitrary complex number.
It is important to mention here that when the operator `{\it a}'
operates on an even
coherent state can generate an odd coherent state with eigenvalue $\alpha$
and different normalization constant.
\begin{equation}
a\mid \alpha_{+}>=\alpha \sqrt{tanh \mid \alpha \mid^{2}}~\mid \alpha_{-}>,
\end{equation}
and similarly
\begin{equation}
a\mid \alpha_{-}>=\alpha \sqrt{coth \mid \alpha \mid^{2}}~\mid \alpha_{+}>.
\end{equation}

The decomposition of the even and
odd coherent states in terms of number states can be obtained by using Eq.(3)
\begin{eqnarray}
\mid \alpha_{+}>&=&N_{+}e^{-\frac{\mid \alpha \mid^{2}}{2}}\sum_{n}
\frac{1+(-1)^{n}}{\sqrt{n!}} \alpha^{n} \mid n>,\nonumber\\
\mid \alpha_{-}>&=&N_{-}e^{-\frac{\mid \alpha \mid^{2}}{2}}\sum_{n}
\frac{1-(-1)^{n}}{\sqrt{n!}} \alpha^{n} \mid n>.\nonumber\\
\end{eqnarray}
It can be easily verify from the above equation that even coherent state can
only be expressed in terms of even number state and odd coherent state in
terms of odd number state.
The photon distribution function can be obtained from Eq.(15)
\begin{eqnarray}
P_{(+)}(n)&=&\left\{ \begin{array}{ll}
\frac{\mid \alpha \mid^{4k}}{cosh\mid \alpha \mid^{2}(2k)!} &
\mbox{for n=2k}\\
0 & \mbox{for n=2k+1}
\end{array} ,\right. \nonumber\\
P_{(-)}(n)&=&\left\{ \begin{array}{ll}
0 & \mbox{for n=2k}\\
\frac{\mid \alpha \mid^{2(2k+1}}{sinh\mid \alpha \mid^{2}(2k+1)!} &
\mbox{for n=2k+1}
\end{array}. \right. \nonumber\\
\end{eqnarray}
The photon distribution function for the even coherent states exhibits the
property that the probability of finding odd
number of photon becomes zero. Similarly the probability of finding the even
number of photons in case of odd coherent states becomes zero. Thus
the probability distribution function for these states is highly oscillatory
function.

The expectation values of the first order moments for the annihilation and
creation operators for the field in even and odd coherent states is zero
which can be verified by using Eqs.(13),(14)
\begin{eqnarray}
<a>_{+}&=&<\alpha_{+} \mid a \mid \alpha_{+}>, \nonumber\\
&=&\alpha \frac{N_{+}}{N_{-}}<\alpha_{+}\mid \alpha_{-}>=0.
\end{eqnarray}
The above result can be obtained because even
and odd coherent states are orthogonal states[1].
The similar result is also true for the odd
coherent states.
Also the expectation values of the second order moments are
\begin{eqnarray}
<a^{2}>_{(\pm)}&=&<\alpha_{\pm} \mid a^{2} \mid \alpha_{\pm}>=\alpha^{2},
\nonumber\\
<a^{\dag}a>_{(+)}&=&<\alpha_{+} \mid a^{\dag}a \mid \alpha_{+}>
=\mid \alpha \mid^{2}tanh\mid \alpha \mid^{2},\nonumber\\
<a^{\dag}a>_{(-)}&=&\mid \alpha \mid^{2}coth\mid \alpha \mid^{2}.
\end{eqnarray}
Then by defining the quadratures of the electromagnetic field mode as
\begin{eqnarray}
X_{1}&=&\frac{a+a^{\dag}}{2},\nonumber\\
X_{2}&=&\frac{a-a^{\dag}}{2i},
\end{eqnarray}
The variances in the two quadratures of even coherent state are
\begin{eqnarray}
\Delta X_{1}^{2}&=&\frac{2\mid \alpha \mid^{2}tanh\mid \alpha \mid^{2}
+2 \mid \alpha \mid^{2} cos 2\theta+1}{4},\nonumber\\
\Delta X_{2}^{2}&=&\frac{2\mid \alpha \mid^{2}tanh\mid \alpha \mid^{2}
-2 \mid \alpha \mid^{2} cos 2\theta+1}{4},
\end{eqnarray}
where we have expressed the complex quantity as $\alpha$=$\mid \alpha \mid
e^{i\theta}$ and $\theta$ be the phase of the coherent state
amplitude. Eq.(20) shows that
for $\theta$=$\pi/2$ the first quadrature shows some amount of squeezing for
small values of $\mid \alpha \mid$ and for $\theta$=0 second quadrature is
squeezed. In Fig.(1) we have plotted $\Delta X_{1}^{2}$ against
$\mid \alpha \mid$, for $\theta$=$\pi/2$. The figure illustrates the region
for squeezing. This nonclassical behavior of even coherent states is also
shown in Refs.[2,3], where  $\alpha$ is considered
to be real and is predicted that squeezing occurs in second quadrature.
We, on the other hand, have predicted the possibility of squeezing
alternatively in both the quadratures depending upon the phase of the
complex amplitude $\mid \alpha \mid$.  For odd coherent states we
get the same expressions as in Eq.(20), but tanh $\mid \alpha \mid^{2}$
 should be replaced by coth $\mid \alpha \mid^{2}$. Also odd coherent states
do not exhibit the property of second-order squeezing.

The variances into the photon number operators for even coherent states
can also be calculated by using Eq.(16).
\begin{eqnarray}
\sigma_{n_{+}}&=&<\alpha_{+}\mid (a^{\dag}a)^{2} \mid \alpha_{+}>
-<\alpha_{+}\mid a^{\dag}a \mid \alpha_{+}>^{2},\nonumber\\
&=&\mid \alpha \mid^{4}+\mid \alpha \mid^{2}tanh\mid \alpha \mid^{2}
-\mid \alpha \mid^{4}tanh~^{2}\mid \alpha \mid^{2},\nonumber\\
\end{eqnarray}
and for odd coherent states
\begin{eqnarray}
\sigma_{n_{-}}&=&<\alpha_{-}\mid (a^{\dag}a)^{2} \mid \alpha_{-}>
-<\alpha_{-}\mid a^{\dag}a \mid \alpha_{-}>^{2},\nonumber\\
&=&\mid \alpha \mid^{4}+\mid \alpha \mid^{2}coth\mid \alpha \mid^{2}
-\mid \alpha \mid^{4}coth~^{2}\mid \alpha \mid^{2}.
\end{eqnarray}
For large values of $\mid \alpha \mid$, the photon number variances for
even and odd coherent states become equal and we have
\begin{equation}
\sigma_{n_{\pm}}=\mid \alpha \mid^{2}.
\end{equation}

\section{multimode even and odd coherent states}

In this section we will discuss the properties of the multimode even and
odd coherent states. We define the multimode even and odd coherent states
as
\begin{equation}
\mid {\bf A_{\pm}}>=N_{\pm} (\mid {\bf A}> \pm \mid -{\bf A}>),
\end{equation}
where the multimode coherent state $\mid {\bf A}>$ is
\begin{equation}
\mid {\bf A}>=\mid \alpha_{1},\alpha_{2},\alpha_{3},......,\alpha_{n}>
=D({\bf A}) \mid {\bf 0}>,\nonumber\\
\end{equation}
and the multimode coherent state is created from multimode vacuum state
$\mid {\bf 0}>$ by the multimode displacement operator $D({\bf A})$ which
is the exact analog of one mode displacement operator (Eq.(2)).
The normalization constants for multimode even and odd coherent states
become
\begin{eqnarray}
N_{+}&=&\frac{e^{\frac{\mid {\bf A} \mid^{2}}{2}}}
{2 \sqrt{cosh\mid {\bf A}\mid^{2}}},\nonumber\\
N_{-}&=&\frac{e^{\frac{\mid {\bf A} \mid^{2}}{2}}}
{2 \sqrt{sinh\mid {\bf A}\mid^{2}}}.
\end{eqnarray}
where {\bf A}=$\alpha_{1},\alpha_{2},\alpha_{3},....,\alpha_{n}$ is a complex
vector and its modulus is
\begin{equation}
\mid {\bf A} \mid^{2}=\mid \alpha_{1} \mid^{2}+\mid \alpha_{2} \mid^{2}
+........+\mid \alpha_{n} \mid^{2}=\sum_{m=1}^{n}\mid \alpha_{m} \mid^{2}.
\nonumber\\
\end{equation}
Such multimode even and odd coherent states can be decomposed into multimode
number states as
\begin{eqnarray}
\mid {\bf A_{+}}>&=&N_{+}\sum_{n}\frac{e^{-\mid {\bf A}\mid^2}
\alpha_{1}^{n_{1}}....\alpha_{n}^{n_{n}}}{\sqrt{n_{1}!}...\sqrt{n_{n}!}}
(1+(-1)^{n_{1}+n_{2}+...n_{n}}) \mid {\bf n}>,\nonumber\\
\mid {\bf A_{-}}>&=&N_{-}\sum_{n}\frac{e^{-\mid {\bf A}\mid^2}
\alpha_{1}^{n_{1}}....\alpha_{n}^{n_{n}}}{\sqrt{n_{1}!}...\sqrt{n_{n}!}}
(1-(-1)^{n_{1}+n_{2}+...n_{n}}) \mid {\bf n}>,\nonumber\\
\end{eqnarray}
where $\mid {\bf n}>=\mid n_{1},n_{2},.....n_{n}>$ being the multimode
number state. Also from Eq.(24), we can derive an important relation for
the multimode even and odd coherent states, i.e.,
\begin{eqnarray}
a_{i} \mid {\bf A_{+}}>&=&\alpha_{i} \sqrt{tanh\mid {\bf A}\mid^{2}}
\mid {\bf A_{-}}>,\nonumber\\
a_{i} \mid {\bf A_{-}}>&=&\alpha_{i} \sqrt{coth\mid {\bf A}\mid^{2}}
\mid {\bf A_{+}}>.\nonumber\\
\end{eqnarray}
{}From above equation it can be easily verified that the expectation
value of the first order moments
of the annihilation and creation operators for the ith mode for the
multimode even and odd coherent states become zero.
The probability of finding {\bf n} photons in multimode even and odd coherent
states can be worked out with the help of Eq.(28).
\begin{eqnarray}
P_{+} (n)&=&\frac{\mid \alpha_{1} \mid^{2n_{1}}
\mid \alpha_{2} \mid^{2n_{2}}...\mid \alpha_{n} \mid^{2n_{n}} }
{(n_{1}!)(n_{2}!)....(n_{n}!)cosh\mid {\bf A}\mid^{2}},
{}~~~{\scriptsize n_{1}+n_{2}+....+n_{n}=2k},\nonumber\\
 P_{-} (n)&=&\frac{\mid \alpha_{1} \mid^{2n_{1}}
\mid \alpha_{2} \mid^{2n_{2}}...\mid \alpha_{n} \mid^{2n_{n}} }
{(n_{1}!)(n_{2}!)....(n_{n}!)sinh\mid {\bf A}\mid^{2}},
{}~~~{\scriptsize n_{1}+n_{2}+....+n_{n}=2k+1}.\nonumber\\
\end{eqnarray}
 Multimode coherent states are the product of independent coherent states
of each mode and photon distribution function is the product of independent
Poissonian distribution functions. But in the present case of multimode
even and odd coherent states we cannot factorize their multimode
photon distribution functions due to the presence of the nonfactorizable
$cosh\mid {\bf A}\mid^{2}$ and $sinh\mid {\bf A}\mid^{2}$.
This fact implies the phenomenon of statistical dependences of different modes
of these states on each other.

In order to describe the properties of the distribution functions from Eq.(30),
 we will calculate the symmetric 2N$\times$2N dispersion matrix
for multimode field's quadrature components. For even and odd coherent states
we have
\begin{equation}
<{\bf A}_{\pm} \mid a_{i}a_{k} \mid {\bf A}_{\pm}>=\alpha_{i}\alpha_{k},
\end{equation}
and complex conjugate values of the above equation for
$<{\bf A}_{\pm} \mid a_{i}^{\dag}a_{k}^{\dag} \mid {\bf A}_{\pm}>$.
Since the quantity $<{\bf A}_{\pm} \mid a_{i} \mid {\bf A}_{\pm}>$ is equal to
zero the above equation gives two N$\times$N blocks of the dispersion matrix
For other two N$\times$N blocks of this matrix ,
we have for the multimode even coherent states
\begin{equation}
\sigma_{(a_{i}^{\dag}a_{k})}^{+}=<{\bf A}_{+} \mid
\frac{1}{2}(a_{i}^{\dag}a_{k}
+a_{k}a_{i}^{\dag})\mid {\bf A_{+}}>=\alpha_{i}^{*}\alpha_{k}
tanh\mid {\bf A} \mid^{2}+\frac{1}{2} \delta_{ik},\nonumber\\
\end{equation}
and for multimode odd coherent states
\begin{equation}
\sigma_{(a_{i}^{\dag}a_{k})}^{-}=<{\bf A}_{-} \mid
\frac{1}{2}(a_{i}^{\dag}a_{k}
+a_{k}a_{i}^{\dag})\mid {\bf A_{-}}>=\alpha_{i}^{*}\alpha_{k}
coth\mid {\bf A} \mid^{2}+\frac{1}{2} \delta_{ik}.\nonumber\\
\end{equation}
For the dispersion matrix, the mean values of the photon
numbers $n_{i}=a_{i}^{\dag}a_{i}$ for multimode even and odd coherent states
are following
\begin{eqnarray}
<{\bf A_{+}}\mid n_{i} \mid {\bf A_{+}}>&=&\mid \alpha_{i}\mid^{2}
tanh\mid {\bf A}\mid^{2},\nonumber\\
<{\bf A_{-}}\mid n_{i} \mid {\bf A_{-}}>&=&\mid \alpha_{i}\mid^{2}
coth\mid {\bf A}\mid^{2}.
\end{eqnarray}
Taking into account the above equation, the symmetric N$\times$N dispersion
matrices for photon number operators can be obtained from the above
given distribution functions for multimode even and odd coherent states.
By defining
\begin{equation}
\sigma_{ik}^{\pm}=<{\bf A}_{\pm}\mid n_{i}n_{k}\mid,
{\bf A}_{\pm}>
\end{equation}
the corresponding  expressions in such states are
\begin{eqnarray}
\sigma_{ik}^{+}&=&\mid \alpha_{i} \mid^{2}\mid \alpha_{k} \mid^{2}
sech^{2}\mid {\bf A} \mid^{2}+\mid \alpha_{i} \mid^{2}tanh
\mid {\bf A} \mid^{2}\delta_{ik},\nonumber\\
\sigma_{ik}^{-}&=&-\mid \alpha_{i} \mid^{2}\mid \alpha_{k} \mid^{2}
cosech^{2}\mid {\bf A} \mid^{2}+\mid \alpha_{i} \mid^{2}coth
\mid {\bf A} \mid^{2}\delta_{ik}.\nonumber\\
\end{eqnarray}
As the nondiagonal matrix elements of the dispersion density matrix
are not equal to zero so we can predict an important feature of multimode
even and odd coherent states that  different modes of these states are
correlated with each other. In other words, as we have mentioned before,
there exist some statistical dependences of different modes on each other.

Another interesting property for the multimode even and odd coherent states
is the Q-function and it can be obtained in the following manner. First of all
the density matrices for the multimode even and odd coherent states are
\begin{equation}
\rho_{\pm}=\mid {\bf A_{\pm}}><{\bf A_{\pm}} \mid,
\end{equation}
then the Q-function can be calculated as
\begin{eqnarray}
 Q_{+}({\bf B},{\bf B}^{*})&=&\frac{1}{\pi}<{\bf B} \mid \rho_{+}
\mid {\bf B}>,\nonumber\\
&=&\frac{4}{\pi} N_{+}^{2}e^{-(\mid {\bf A}\mid ^{2}
+\mid {\bf B}\mid ^{2})}\mid
cosh({\bf A {\bf B}^{*}})\mid ^{2},\nonumber\\
Q_{-}({\bf B},{\bf B}^{*})&=&\frac{1}{\pi}<{\bf B} \mid \rho_{-}
\mid {\bf B}>,\nonumber\\
&=&\frac{4}{\pi} N_{-}^{2}e^{-(\mid {\bf A}\mid ^{2}
+\mid {\bf B}\mid ^{2})}\mid sinh({\bf A
{\bf B}^{*}})\mid ^{2},
\end{eqnarray}
where $\mid {\bf B}>=\mid \beta_{1},\beta_{2},.....,\beta_{n}>$
is another multimode coherent state with multimode eigenvalue
${\bf B}=(\beta_{1},\beta_{2},.....,\beta_{n})$.
We call these functions for even and odd coherent states
as the Q-functions for the Schrodinger cat states. In Fig.(2) we have plotted
a three-dimensional plots of Q-function
for single-mode even coherent states and
for different choices of the quantity $\mid \alpha \mid$
versus $Re (\beta)$ and $Im (\beta)$. Plots illustrate the fact that for small
values of $\mid \alpha \mid$, we get a single peak of the probability function
and how this single peak begins to split into two peaks for the increasing
values of $\mid \alpha \mid$.
Figs.(3) show Q-function for single-mode odd coherent state and it shows the
crater type behaviour for the Q-function for small values of the
quantity $\mid \alpha \mid$ and for its larger values the Q-function begins to
split into two peaks in a similar manners as in case of even coherent states.
Hence we have shown that the behaviour of the plots for Q-functions of even
and odd coherent states are essentially different for small amplitudes
of $\mid \alpha \mid$ and become very similar for its large values.

The Wigner function for the multimode coherent states is [19]
\begin{equation}
W_{{\bf A,B}}({\bf q,p})=2^{N}exp[-2{\bf Z Z^{*}}+2{\bf A Z^{*}}+2{\bf BZ^{*}}
-{\bf AB^{*}}-\frac{\mid {\bf A}\mid^{2}}{2}-\frac{\mid {\bf B}\mid^{2}}{2}],
\nonumber\\
\end{equation}
where
\begin{equation}
{\bf Z}=\frac{{\bf q+i p}}{\sqrt{2}}.
\end{equation}
For even and odd coherent states the Wigner  function will be
the following different combinations of the above equation
\begin{eqnarray}
W_{{\bf A}_{+}}({\bf q,p})&=&\mid N_{+}\mid^{2}[W_{{\bf (A,B=A)}}({\bf q,p})+
W_{{\bf (A,B=-A)}}({\bf q,p}),\nonumber\\
&&+W_{{\bf (-A,B=A)}}({\bf q,p})+W_{({\bf -A,B=-A)}}({\bf q,p})],\nonumber\\
W_{{\bf A_{-}}}({\bf q,p})&=&\mid N_{-} \mid^{2}[ W_{{\bf (A,B=A)}}({\bf q,p})-
W_{{\bf (A,B=-A)}}({\bf q,p}),\nonumber\\
&&-W_{{\bf (-A,B=A)}}({\bf q,p})+ W_{{\bf (-A,B=-A)}}({\bf q,p})],\nonumber\\
\end{eqnarray}
where the explicit forms of $N_{\pm}$ are given in Eq.(26).
For multimode case we use the following notations
\begin{eqnarray}
{\bf AZ^{*}}&=&\alpha_{1}Z_{1}^{*}+\alpha_{2}Z_{2}^{*}+....\alpha_{n}Z_{n}^{*},
\nonumber\\
{\bf ZZ^{*}}&=&Z_{1}Z_{1}^{*}+Z_{2}Z_{2}^{*}+.....+Z_{n}Z_{n}^{*}.
\end{eqnarray}
After describing some properties of the multimode even and odd coherent states,
now in the following subsection we will discuss the case of two-mode even and
odd coherent states.

\subsection{Two mode even and odd coherent states}

For two mode even and odd coherent states, Eq.(24) can be redefined as
\begin{equation}
\mid {\bf A}_{\pm}>=N_{\pm} (\mid \alpha_{1}+\alpha_{2}>
\pm \mid -\alpha_{1}-\alpha_{2}>,
\end{equation}
and the normalization constants becomes
\begin{eqnarray}
{\bf N}_{+}&=&\frac{e^{\frac{\mid \alpha_{1}\mid^{2}+\mid \alpha_{2}
\mid^{2}}{2}}}{2\sqrt{cosh(\mid \alpha_{1}\mid^{2}+\mid \alpha_{2}\mid^{2})}},
\nonumber\\
{\bf N}_{-}&=&\frac{e^{\frac{\mid \alpha_{1}\mid^{2}+\mid \alpha_{2}
\mid^{2}}{2}}}{2\sqrt{sinh(\mid \alpha_{1}\mid^{2}+\mid \alpha_{2}\mid^{2})}}.
\end{eqnarray}
Also from Eq.(29) we can define the relation
\begin{eqnarray}
a_{i}\mid {\bf A}_{+}>&=&\alpha_{i}\sqrt{tanh((\mid \alpha_{1}\mid^{2}+\mid
\alpha_{2}\mid^{2})}\mid {\bf A}_{-}>,\nonumber\\
a_{i}\mid {\bf A}_{-}>&=&\alpha_{i}\sqrt{coth((\mid \alpha_{1}\mid^{2}+\mid
\alpha_{2}\mid^{2})}\mid {\bf A}_{+}>,~~~i=1,2.\nonumber\\
\end{eqnarray}
Above relation is an important result and with the help of this expression
it is very simple to evaluate the first and higher order expectation values
of different operators. For instance for even coherent states the first and
second order moments are
\begin{eqnarray}
<a_{i}>_{+}&=&<{\bf A}_{+}\mid a_{i} \mid {\bf A}_{+}>,\nonumber\\
&=&\frac{\alpha_{i}}{4e^{-(\mid \alpha_{1} \mid^{2}+\mid \alpha_{2} \mid^{2})}
sinh(\mid \alpha_{1} \mid^{2}+\mid \alpha_{2} \mid^{2})}
<{\bf A}_{+}\mid {\bf A}_{-}>=0,\nonumber\\
\end{eqnarray}
as even and odd coherent states are orthogonal states. Also
\begin{eqnarray}
<a_{i}^{2}>_{+}&=&\alpha_{i}^{2}<{\bf A}_{+} \mid {\bf A}_{+}>=\alpha_{i}^{2},
\nonumber\\
<a_{i}^{\dag}a_{i}>&=&\mid \alpha_{i}\mid^{2}\frac{\mid {\bf N}_{+}\mid^{2}}
{\mid{\bf N}_{-}\mid^{2}}<{\bf A}_{+} \mid {\bf A}_{+}>,\nonumber\\
&=&\mid \alpha_{i}\mid^{2}tanh(\mid \alpha_{1} \mid^{2}+
\mid \alpha_{2} \mid^{2}),\nonumber\\
for~~ i=1,2\nonumber\\
<a_{1}^{\dag}a_{2}>_{+}&=&\alpha_{1}^{*}\alpha_{j}
tanh(\mid \alpha_{1} \mid^{2}+\mid \alpha_{2} \mid^{2}).
\end{eqnarray}
Such expectation values of first and second order moments will be used to
evaluate squeezing in two mode even coherent states

For two mode even coherent states we may define the quadratures of the
field modes as
\begin{eqnarray}
d_{1}&=&\frac{d+d^{\dag}}{2},\nonumber\\
d_{2}&=&\frac{d-d^{\dag}}{2i},
\end{eqnarray}
where d is the superposition of the two field modes
\begin{equation}
d=\frac{a_{1}+a_{2}}{\sqrt{2}}.
\end{equation}
The variances into the two quadratures are
\begin{equation}
\Delta d_{1,2}^{2}=<(d_{1}\pm d_{2})^{2}>-<d_{1}\pm d_{2}>^{2},
\end{equation}
where `+' sign corresponds to $\Delta d_{1}$ and `-' for $\Delta d_{2}$.
After substituting the values of second order moments from Eq.(47) into Eq.(50)
we have
\begin{eqnarray}
\Delta d_{1,2}^{2}&=&\frac{1}{4}\left[ (\mid \alpha_{1} \mid^{2}+
\mid \alpha_{2} \mid^{2}
+2 \mid \alpha_{1} \mid \mid \alpha_{2} \mid cos(\theta_{1}-\theta_{2}))\right.
\nonumber\\
&& \times tanh (\mid \alpha_{1} \mid^{2}+ \mid \alpha_{2} \mid)
 \pm( \mid \alpha_{1} \mid^{2}\nonumber\\
&&+ \left. \mid \alpha_{2} \mid^{2}
+2 \mid \alpha_{1} \mid \mid \alpha_{2} \mid
cos(\theta_{1}+\theta_{2}))\right],
\end{eqnarray}
where we have defined the complex quantity as $\alpha_{i}= \mid \alpha_{i} \mid
e^{i\theta_{i}}$.
In Fig.(4), we have plotted such variance in the second quadrature versus
$\mid \alpha_{1}\mid$ and $\mid \alpha_{2}\mid$, for $\theta_{1}=\theta_{2}=0$
, plot shows the region where we can get maximum amount of squeezing in
two-mode even coherent states.

The photon distribution function gives the probability of finding
$2k$ photons for two-mode even coherent state and is defined as
\begin{equation}
P_{+}(2k)=\frac{(\mid \alpha_{1} \mid^{2}+\mid \alpha_{2} \mid^{2})^{2k}}
{(2k)!cosh(\mid \alpha_{1} \mid^{2}+\mid \alpha_{2} \mid^{2})},
\end{equation}
where $2k=n_{1}+n_{2}$, for both $n_{1}$ and $n_{2}$ to be even or odd numbers.
 Similarly for two-mode odd coherent state it gives the probability
of finding 2k+1 photons
\begin{equation}
P_{-}(2k+1)=\frac{(\mid \alpha_{1} \mid^{2}+\mid \alpha_{2} \mid^{2})^{(2k+1)}}
{(2k+1)!sinh(\mid \alpha_{1} \mid^{2}+\mid \alpha_{2} \mid^{2})}.
\end{equation}
For this case $n_{1}$ is  even and $n_{2}$ is odd number or vise versa.
The dispersion matrices for the number operators in case of two-mode even
and odd coherent states may be defined in terms of 2$\times$2 matrices as
\begin{eqnarray}
\sigma_{n_{1},n_{2}}^{\pm}&=&\left( \begin{array}{clcr}
\sigma_{11}^{\pm} &  \sigma_{12}^{\pm}\\
\sigma_{21}^{\pm} &  \sigma_{22}^{\pm}
\end{array} \right), \nonumber\\
\end{eqnarray}
where
\begin{eqnarray}
\sigma_{11}^{+}&=&\mid \alpha_{1}\mid^{4}sech^{2}(\mid \alpha_{1}\mid^{2}+
\mid \alpha_{2}\mid^{2})
+\mid \alpha_{1}\mid^{2}tanh(\mid \alpha_{1}\mid^{2}+\mid \alpha_{2}\mid^{2}),
\nonumber\\
\sigma_{12}^{+}&=&\mid \alpha_{1}\mid^{2}\mid \alpha_{2}\mid^{2} sech^{2}
(\mid \alpha_{1}\mid^{2}+\mid \alpha_{2}\mid^{2}),\nonumber\\
\sigma_{11}^{-}&=&
\mid \alpha_{1}\mid^{4}cosech^{2}(\mid \alpha_{1}\mid^{2}+
\mid \alpha_{2}\mid^{2})
+\mid \alpha_{1}\mid^{2}coth(\mid \alpha_{1}\mid^{2}+\mid \alpha_{2}\mid^{2}),
\nonumber\\
\sigma_{12}^{-}&=&-\mid \alpha_{1}\mid^{2}\mid \alpha_{2}\mid^{2} cosech^{2}
(\mid \alpha_{1}\mid^{2}+\mid \alpha_{2}\mid^{2}).
\end{eqnarray}
For single-mode case it has been shown [2] that the photon distribution
functions  demonstrate super and sub-Poissonian properties for even and odd
coherent states respectively. It can be easily shown on the basis of the above
expressions that same conclusions may be drawn for the two-mode (and
multimode) even and odd coherent states.
In order to show the correlations between these two modes of even and odd
coherent states, we define the correlation coefficients as
\begin{equation}
R_{\pm}=\frac{\sigma_{12}^{\pm}}{\sqrt{\sigma_{11}^{\pm}\sigma_{22}^{\pm}}},
\end{equation}
and by using Eq.(55), we obtain the expression
\begin{equation}
R_{\pm}=\frac{\pm\mid\alpha_{1}\mid\mid\alpha_{2}\mid}
{\sqrt{(\mid\alpha_{1}\mid^{2}\pm\frac{1}{2}sinh2(\mid\alpha_{1}\mid^{2}+
\mid\alpha_{2}\mid^{2})(\mid\alpha_{2}\mid^{2}\pm\frac{1}{2}
sinh2(\mid\alpha_{1}\mid^{2}+
\mid\alpha_{2}\mid^{2})}}.\nonumber\\
\end{equation}
It is clear from the above equation that for large values of
$\mid\alpha_{1}\mid,~\mid\alpha_{2}\mid$ this correlation tends to zero, but
for small values of these quantities there exists an essential correlations
between these two modes of even and odd coherent states.
In Fig.(5) we have plotted $R_{+}$ versus $\mid\alpha_{1}\mid$ for different
values of $\mid\alpha_{2}\mid$. It is also evident from the graph that
the correlations between two-mode even coherent states occur for small values
of $\mid\alpha_{1}\mid$ and $\mid\alpha_{2}\mid$ and they disappear for large
values. For such small values these correlations
 are also responsible for two-mode squeezing as shown in Fig.(4).

In conclusions, we would emphasize that the suggested multimode
generalization of even and odd coherent states might be extended for other
superpositions of single-mode
coherent states  discussed by Schleich et al. [20,21]. These
states are superpositions of coherent states with arbitrary relative phase
factors. The photon distributions for such multimode states  should demonstrate
 essential dependences of correlations between different
modes on the phase factors. Such dependences produce the structure of the
photon distribution functions similar to the structure in case of two-mode
squeezed vacuum state [22], with oscillatory behaviour like for single-mode
squeezed light, obtained by Schleich and Wheeler [23].

\newpage

{\bf Acknowledgements}\\

The authors wish to thanks Prof. S. Solimeno and Prof. F. Zaccaria for some
useful discussions. The research of N.A.A. was supported by the
International Center for Theoretical Physics Programme for Research and
Training in Italian Laboratories. V.I.M. wishes to thank the Dipartimento
di Scienze Fisiche, Universita' di Napoli and I.N.F.N. for the kind
hospitality.

\newpage

{\bf References}\\

\begin{enumerate}

\item V.V. Dodonov, I.A. Malkin and V.I. Man'ko, Physica {\bf 72}, 579 (1974).

\item V. Buzik, A. VIdiella-Barranco and P.L. Knight, Phys. Rev. A {\bf 45},
6750 (1992).

\item Y. Xia and G. Guo, Phys. Lett. A {\bf 136}, 281 (1989).

\item M. Hillery, Phys. Rev. A {\bf 36} 3796 (1987).

\item C.C. Gerry, Opt. Commun. {\bf 91}, 247 (1992).

\item C.C. Gerry and E.E. Hach III, Phys. Lett A {\bf 74}, 185 (1993).

\item J. Gea-Banacloche, Phys. Rev. Lett. {\bf 65}, 3385 (1990); Phys. Rev. A
{\bf 44}, 5913 (1991).

\item V. Buzik, H. Moya-Cessa, P.L. Knight and J.J.D. Phoenix, Phys. Rev. A
{\bf 45}, 8190 (1992).

\item M. Brune, S. Haroche, J.M Raimond, L. Davidovich and N. Zagury,
Phys. Rev. A {\bf 45}, 5193 (1992).

\item N.A. Ansari, L. Di Fiore, M.A. Man'ko, V.I. Man'ko, S. Solimeno
and F. Zaccaria, Phys. Rev. A (submitted).

\item D. Stoler, Phys. Rev. D {\bf 4}, 2309 (1971).

\item J.N. Hollenhorst, Phys. Rev. D {\bf 19}, 1669 (1979).

\item H.P. Yuen, Phys. Rev. A {\bf 13}, 2226 (1976).

\item For review on squeezed states see, D.F. Walls, Nature {\bf 306},
141 (1983).

\item V.V. Dodonov, E.V. Kurmyshev and V.I. Man'ko, Phys. Lett. A {\bf 79},
150 (1980).

\item N.A. Ansari, Opt. Lett. (submitted).

\item R.J. Glauber, Phys. Rev. Lett. {\bf 10}, 84 (1963).

\item See for example,
I.A. Malkin and V.I. Man'ko, "Dynamical symmetries and coherent states of
quantum systems", (Nauka publishers, Mascow, 1979) [in Russian].

\item V.V. Dodonov and V.I. Man'ko in "Invariants and evolution of
nonstationary
 quantum systems", Proceedings of Lebedev Physics institute {\bf 183},
ed. by M.A. Markov, (Nova Science Publishers, Commack, N.Y. 1989).

\item W. Schleich, M. Pernigo and Fam Le Kien, Phys. Rev. A {\bf 44},
2174 (1991).

\item W. Schleich, J.P. Dowling. R.J. Horowicz and S. Varro, in New
Frontiers in Quantum Optics and Quantum Electrodynamics, ed. by A. Barut
(Plenum, N.Y., 1990).

\item G. Schrade, V.M. Akulin, V.I. Man'ko and W. Schleich, Phys. Rev A
(in press).

\item W. Schleich and Wheeler, J. Opt. Soc. Am B {\bf 4}, 1715 (1987).

\end{enumerate}

\newpage

{\bf Figure Captions}\\

\begin{description}

\item[Fig.(1).] Variance in the first quadrature of the field mode
$\Delta X_{1}^{2}$ versus $\mid \alpha \mid$ for $\theta=\pi/2$.

\item[Fig.(2a).] Three dimensional plot of Q-function for single-
mode even coherent light $Q_{+}(\beta,\beta^{*})$ versus Re($\beta$) and
Im($\beta$) for $\mid \alpha \mid$=0.5.

\item[Fig.(2b).] $Q_{+}(\beta,\beta^{*})$ versus Re($\beta$) and
Im($\beta$) for $\mid \alpha \mid$=1.3.

\item[Fig.(2c).] $Q_{+}(\beta,\beta^{*})$ versus Re($\beta$) and
Im($\beta$) for $\mid \alpha \mid$=3.

\item[Fig.(3a).] Three dimensional plot of Q-function for single-
mode even coherent light $Q_{-}(\beta,\beta^{*})$ versus Re($\beta$) and
Im($\beta$) for $\mid \alpha \mid$=0.5.

\item[Fig.(3b).] $Q_{-}(\beta,\beta^{*})$ versus Re($\beta$) and
Im($\beta$) for $\mid \alpha \mid$=1.3.

\item[Fig.(3c).] $Q_{-}(\beta,\beta^{*})$ versus Re($\beta$) and
Im($\beta$) for $\mid \alpha \mid$=3.

\item[Fig.(4).] Three dimensional plot of the variance in the second quadrature
of the field modes in case of two-mode even coherent states $\Delta d_{2}^{2}$
versus $\mid \alpha_{1}\mid$ and $\mid \alpha_{2}\mid$, for $\theta_{1}=
\theta_{2}=0$.

\item[Fig.(5)] Correlation coefficient $R_{+}$ for two-mode photon numbers of
even coherent states versus  $\mid \alpha_{1}\mid$ for  $\mid \alpha_{2}\mid$
=0.5, 1 and 1.5.

\end{description}

\end{document}